\documentclass[12pt]{article}
\usepackage{graphicx}

\def \bea{\begin{eqnarray}}
\def \beq{\begin{equation}}
\def \ca{{\cal A}}
\def \cb{{\cal B}}
\def \eea{\end{eqnarray}}
\def \eeq{\end{equation}}

\def \ol{\overline}
\def \s{\sqrt{2}}
\def \st{\sqrt{3}}
\def \sx{\sqrt{6}}
\def \ta{\tilde{A}}
\def \tc{\tilde{C}}
\def \te{\tilde{E}}
\def \ttl{\tilde{T}}
\def \p{\prime}
\def \im{{\it{i}}}
\def \thet{\theta_\eta}

\begin{document}
\rightline{EFI 09-32}
\rightline{arXiv:0911.2812}
\rightline{November 2009}
\bigskip
\centerline{\bf CHARMED MESON DECAYS TO TWO PSEUDOSCALARS}
\bigskip

\centerline{Bhubanjyoti Bhattacharya\footnote{bhujyo@uchicago.edu} and
Jonathan L. Rosner\footnote{rosner@hep.uchicago.edu}}
\centerline{\it Enrico Fermi Institute and Department of Physics}
\centerline{\it University of Chicago, 5640 S. Ellis Avenue, Chicago, IL 60637}

\begin{quote}

A recent update of data on the decays of $D^0$, $D^+$, and $D_s^+$ to pairs of
light pseudoscalars calls for a renewed analysis of key decay amplitudes and
tests of flavor symmetry. The present data change our previous understanding of
relative phases between amplitudes that describe Cabibbo-favored decays of the
charmed mesons.  The new data also seem to favor a smaller octet-singlet mixing
angle for the $\eta$ and $\eta'$ mesons when singly-Cabibbo-suppressed
processes are taken into account.
We also discuss the effects of the new data on interference between
Cabibbo-favored and doubly-Cabibbo-suppressed decays.

\end{quote}

\section{INTRODUCTION}

The CLEO collaboration have recently reported new results on charmed-meson
decay rates and branching ratios \cite{Mendez:2009}, many with experimental
errors less than present world averages \cite{Amsler:2008}.  This calls for an
update of a previous work \cite{Bhattacharya:2008ss} on extraction of
flavor-topology amplitudes and relative phases between them.  SU(3) flavor
symmetry, applied here, has been shown useful in finding relative strong phases
of amplitudes in $D \to PP$ decays, where $P$ represents a light pseudoscalar
meson \cite{Chau:1983, Chau:1986, Rosner:1999}.

The diagrammatic approach is once again reviewed in Sec.\ II. In Sec.\ III we
talk about Cabibbo-favored decays and a new and previously unexpected relative
phase between two of the amplitudes in light of the new data. We also perform
an analysis of the Cabibbo-favored decay amplitudes using a variable angle for
the octet-singlet mixing in $\eta$ and $\eta'$. Sec.\ IV updates the
singly-Cabibbo-suppressed amplitudes and their role in determining amplitudes
associated with disconnected diagrams.  In Sec.\ V we discuss
doubly-Cabibbo-suppressed decays. We conclude in Sec.\ VI.

\section{DIAGRAMMATIC AMPLITUDE EXPANSION}

The flavor-topology technique for analyzing charmed meson decays makes use of
$SU(3)$ invariant amplitudes. The key amplitudes that describe the physics of
Cabibbo-favored decays have been defined in Ref.\ \cite{Bhattacharya:2008ss},
and include a color-favored tree ($T$), a color-suppressed tree ($C$), an
exchange ($E$), and an annihilation ($A$) amplitude. The Cabibbo-favored
(CF) amplitudes are proportional to the product $V_{ud}V_{cs}$ of the
Cabibbo-Kobayashi-Maskawa (CKM) matrix elements, the singly-Cabibbo-suppressed
amplitudes are proportional to $V_{us}V_{cs}$ or $V_{ud}V_{cd}$, and the
doubly Cabibbo-suppressed amplitudes are proportional to $V_{us}V_{cd}$. We
denote the Cabibbo-favored, singly-Cabibbo-suppressed and doubly-%
Cabibbo-suppressed amplitudes by unprimed, primed, and quantities with a tilde,
respectively.  The relative hierarchy of these amplitudes in terms of the
Wolfenstein parameter $\lambda = \tan\theta_C = 0.2317$ \cite{Amsler:2008} is
$1:\lambda:-\lambda:-\lambda^2$, where $\theta_C$ is the Cabibbo angle.

\section{CABIBBO-FAVORED DECAYS}

Cabibbo-favored $D$ decays have been discussed at length in Refs.\
\cite{Bhattacharya:2008ss, Rosner:1999}, where the $C$, $E$ and $A$ amplitudes
were found to have large phases relative to the dominant $T$
amplitude. In particular, we found \cite{Bhattacharya:2008ss} that the
amplitudes $E$ and $A$ had a relative phase of approximately $180^\circ$. This
conclusion changes when we make use of new branching ratios for Cabibbo-favored
$D$ decays \cite{Mendez:2009}. The phases of $C$ and $E$ relative to $T$ are
unchanged, but in the favored solution $E$ and $A$ now are no
longer separated by a $180^\circ$ phase difference, but are much closer to one
another in phase.  The magnitude of $A$ is smaller than was found in Ref.\
\cite{Bhattacharya:2008ss} and hence the ratio between $|A|$ and $|E|$ is also
smaller: $|A| = (0.21\pm0.09)|E|.$

\begin{table}
\caption{Branching ratios and invariant amplitudes for Cabibbo-favored
decays of charmed mesons to a pair of pseudoscalars.  Here an octet-singlet
mixing angle of $\theta_\eta = \arcsin(1/3) = 19.5^\circ$ has been assumed.
\label{tab:CF1}}
\begin{center}
\begin{tabular}{c l c c c c c}
\hline \hline
Meson & Decay & $\cb$ \cite{Mendez:2009} & $p^*$ & $|{\cal A}|$ & Rep.\
 & Predicted $\cb$ \\
 & mode  & ($\%$) & (MeV) & $(10^{-6} GeV)$ &  & ($\%$)  \\ \hline
$D^0$ &$K^- \pi^+$ &$3.891\pm0.077$ & 861.1 &$2.52\pm0.03$ & $T+E$ & 3.905 \\
       &$\ol{K}^0 \pi^0$  &$2.380\pm0.092$  & 860.4 &$1.97\pm0.04$ & $(C-E)/\s$
 & 2.347 \\
       &$\ol{K}^0 \eta$ &$0.962\pm0.060$  & 771.9 &$1.32\pm0.04$   & $C/\st$
 & 1.002 \\
       &$\ol{K}^0 \eta^\p$&$1.900\pm0.108$ & 564.9 &$2.17\pm0.06$
 & $-(C+3E)/\sx$ & 1.920 \\ \hline
$D^+$ &$\ol{K}^0 \pi^+$  &$3.074\pm0.097$ & 862.4 &$1.41\pm0.02$  & $C+T$
 & 3.090   \\ \hline
$D^+_s$&$\ol{K}^0 K^+$ &$2.98\pm0.17$ & 850.3 &$2.12\pm0.06$ & $C+A$
 & 2.939   \\
 &$\pi^+ \eta$ &$1.84\pm0.15$ & 902.3 &$1.62\pm0.07$ &$(T-2A)/\st$& 1.810 \\
 &$\pi^+ \eta^\p$ &$3.95\pm0.34$ & 743.2 &$2.61\pm0.11$ &$2(T+A)/\sx$
 & 3.603   \\ \hline \hline
\end{tabular}
\end{center}
\end{table}

\begin{figure}
\mbox{\includegraphics[width=0.48\textwidth]{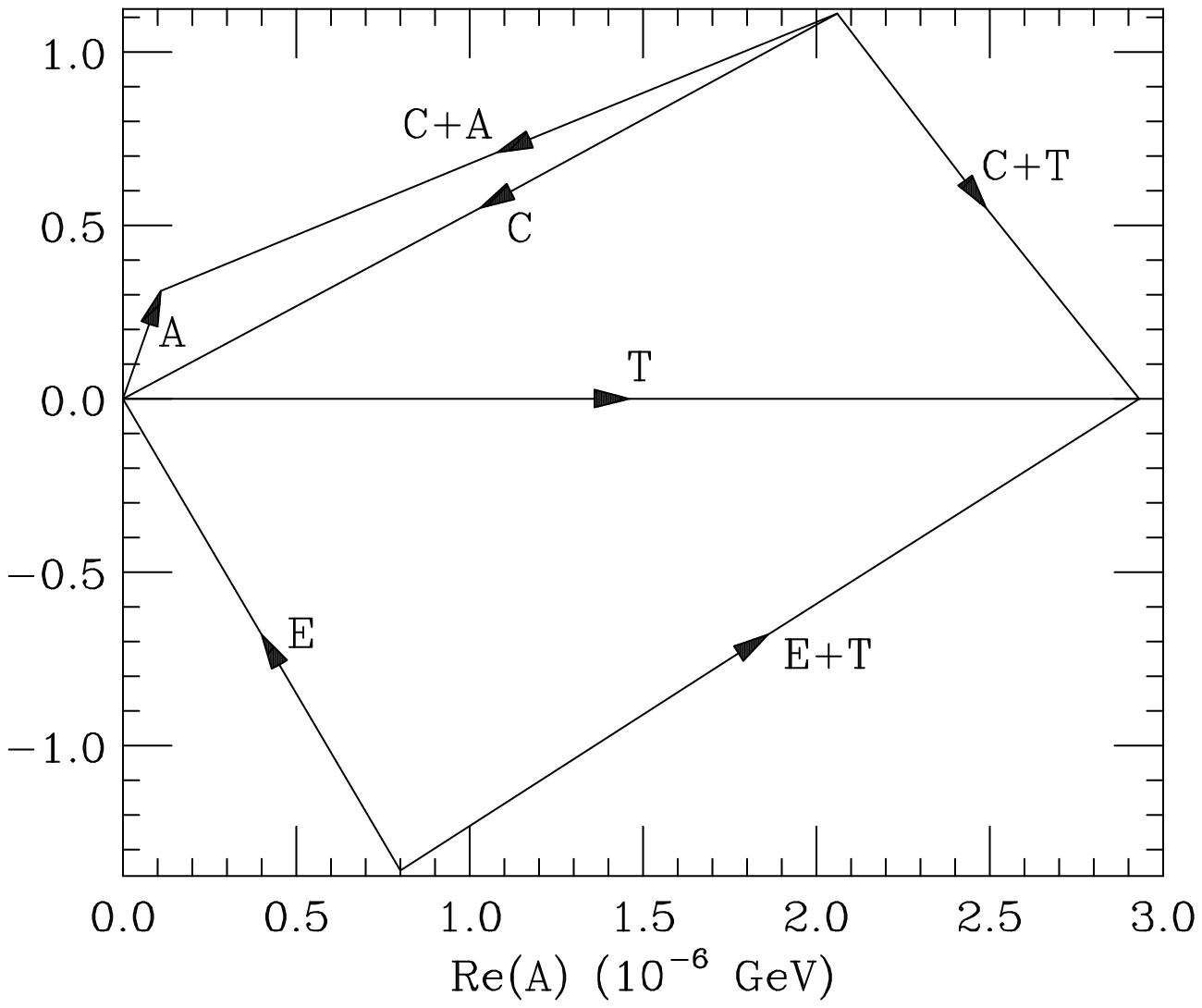}
      \includegraphics[width=0.48\textwidth]{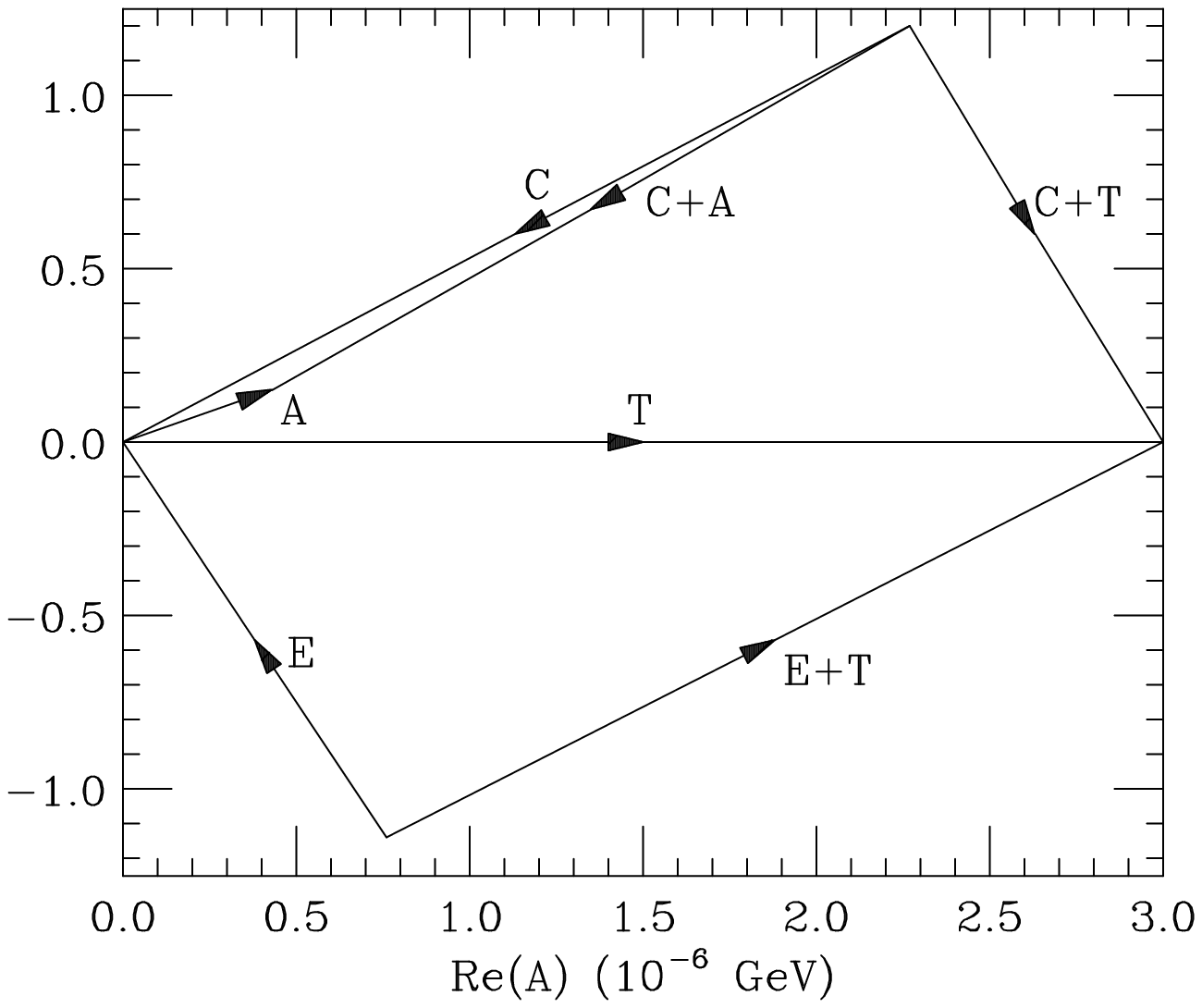}}
\caption{Construction of Cabibbo-favored amplitudes from observed processes
using a least $\chi^2$-fit. The sides $C+T$, $C+A$, and $E+T$ correspond to
measured processes; the magnitudes of other amplitudes listed in Table
\ref{tab:CF1} are also needed to specify $T$, $C$, $E$, and $A$. These
figures correspond to the $|T| > |C|$ solution.  Left: $\thet$ fixed at
$\arcsin(1/3) = 19.5^\circ$ with $\chi^2 = 1.79$ for one degree of freedom.
Right: exact solution with $\thet = 11.7^\circ$ and $\chi^2=0$.
\label{fig:CFA1}}
\end{figure}

In Table \ref{tab:CF1} we list the branching ratios $\cb$ \cite{Mendez:2009}
corresponding to the Cabibbo-favored decay modes and the amplitudes extracted
using $\ca = M_D\left[8\pi\cb\hbar/(p^*\tau)\right]^{1/2}$, where $M_D$ is the
mass of the decaying meson, $\tau$ is its lifetime and $p^*$ is the
center-of-mass 3-momentum of a final state pseudoscalar meson.  We have
described octet-singlet mixing in the $\eta$ and $\eta'$ mesons in terms of
an angle $\theta_\eta$:
\beq
\eta  = - \eta_8 \cos \theta_\eta - \eta_1 \sin \theta_\eta~,~~
\eta' = - \eta_8 \sin \theta_\eta + \eta_1 \cos \theta_\eta~,~~{\rm where}
\label{eqn:a1}
\eeq
\beq
\eta_8 \equiv (u \bar u + d \bar d - 2 s \bar s)/\sqrt{6}~,~~
\eta_1 \equiv (u \bar u + d \bar d + s \bar s)/\sqrt{3}~,
\label{eqn:a2}
\eeq
and have taken $\theta_\eta = \arcsin(1/3) = 19.5^\circ$ so that \cite{etamix}
\beq
\eta  = (s \bar s - u \bar u - d \bar d)/\sqrt{3}~,~~
\eta' = (2 s \bar s + u \bar u + d \bar d)/\sqrt{6}~.
\label{eqn:b}
\eeq
We can extract
$T$, $C$, $E$, and $A$ uniquely (up to a complex conjugation) by defining $T$
to be purely real. The extracted amplitudes, in units of $10^{-6}$ GeV, are:
\bea
T &=&  2.927\pm0.022 \\
C &=& (2.337\pm0.027)\,\exp\left[\im(-151.66\pm0.63)^\circ\right] \\
E &=& (1.573\pm0.032)\,\exp\left[\im( 120.56\pm1.03)^\circ\right] \\
A &=& (0.33\pm0.14)\,\exp\left[\im(  70.47\pm10.90)^\circ\right]
\label{eqn:c}
\eea
These amplitudes are shown on an Argand diagram in the left-hand panel of
Fig.\ \ref{fig:CFA1}.  They were extracted by a least $\chi^2$-fit to the data,
resulting in $\chi^2 = 1.79$ for 1 degree of freedom, and update those quoted
in Refs.\ \cite{Bhattacharya:2008ss, Rosner:1999}.

While the above analysis works fairly well and yields a low value of $\chi^2$,
one could obtain an exact solution ($\chi^2=0$) by introducing one more
parameter; the number of constraints (known branching ratios) would then
equal the number of unknown variables. There are several possible sources of
an extra parameter.  One might add singlet amplitudes, but they are expected to
be much smaller than the non-singlet ones and would result in too large a
parameter space.  A plausible new parameter is the angle $\theta_\eta$
describing octet-singlet mixing in $\eta$ and $\eta'$, which was fixed in the
above analysis to \cite{etamix} $\thet = 19.5^\circ$.  An exact solution is
obtained in this case for $\theta_\eta = 11.7^\circ$. The corresponding
amplitudes, in units of $10^{-6}$ GeV, have been listed below and plotted on
an Argand diagram in the right-hand panel of Fig.\ \ref{fig:CFA1}:
\bea
T &=&  3.003\pm0.023 \\
C &=& (2.565\pm0.030)\,\exp\left[\im(-152.11\pm0.57)^\circ\right] \\
E &=& (1.372\pm0.036)\,\exp\left[\im( 123.62\pm1.25)^\circ\right] \\
A &=& (0.452\pm0.058)\,\exp\left[\im(  19^{+15}_{-14})^\circ\right]
\eea

\begin{table}
\caption{Branching ratios and invariant amplitudes for Cabibbo-favored
decays of charmed mesons to a pair of pseudoscalars with 2 different values of
$\thet$. ($\phi_1 = 45^\circ - \frac{\phi_2}{2}$ and $\phi_2 = 19.5^\circ$.)
\label{tab:CF2}}
\begin{center}
\begin{tabular}{c l c c c c}
\hline \hline
Meson & Decay & $\cb$ \cite{Mendez:2009} & Rep.\ & \multicolumn{2}{c}{Predicted $\cb$ ($\%$)} \\
      & mode  & ($\%$) &  & $\theta_\eta = 11.7^\circ$ & $\theta_\eta
 = 19.5^\circ$    \\ \hline
$D^0$  &$K^- \pi^+$       &3.891$\pm$0.077&  $T+E$       &3.891&3.905\\
       &$\ol{K}^0 \pi^0$  &2.380$\pm$0.092& $(C-E)/\s$   &2.380&2.347\\
       &$\ol{K}^0 \eta$   &0.962$\pm$0.060&~$\frac{C}{\s}\sin(\thet+\phi_1)
       -\frac{\st E}{\s}\cos(\thet+2\phi_1)$&0.962&1.002\\
       &$\ol{K}^0 \eta^\p$&1.900$\pm$0.108&-$\frac{C}{\s}\cos(\thet+\phi_1)
       -\frac{\st E}{\s}\sin(\thet+2\phi_1)$&1.900&1.920\\\hline
$D^+$  &$\ol{K}^0 \pi^+$  &3.074$\pm$0.097& $C+T$        &3.074&3.090\\ \hline
$D^+_s$&$\ol{K}^0 K^+$    &2.98$\pm$0.17  & $C+A$        &2.980&2.939\\
       &$\pi^+ \eta$      &1.84$\pm$0.15  & $T\cos(\thet+\phi_1)
       -\s A\sin(\thet+\phi_1)$ &1.840&1.810\\
       &$\pi^+ \eta^\p$   &3.95$\pm$0.34  & $T\sin(\thet+\phi_1)
       +\s A\cos(\thet+\phi_1)$ &3.950&3.603\\ \hline \hline
\end{tabular}
\end{center}
\end{table}

\begin{figure}
\mbox{\includegraphics[width=0.97\textwidth]{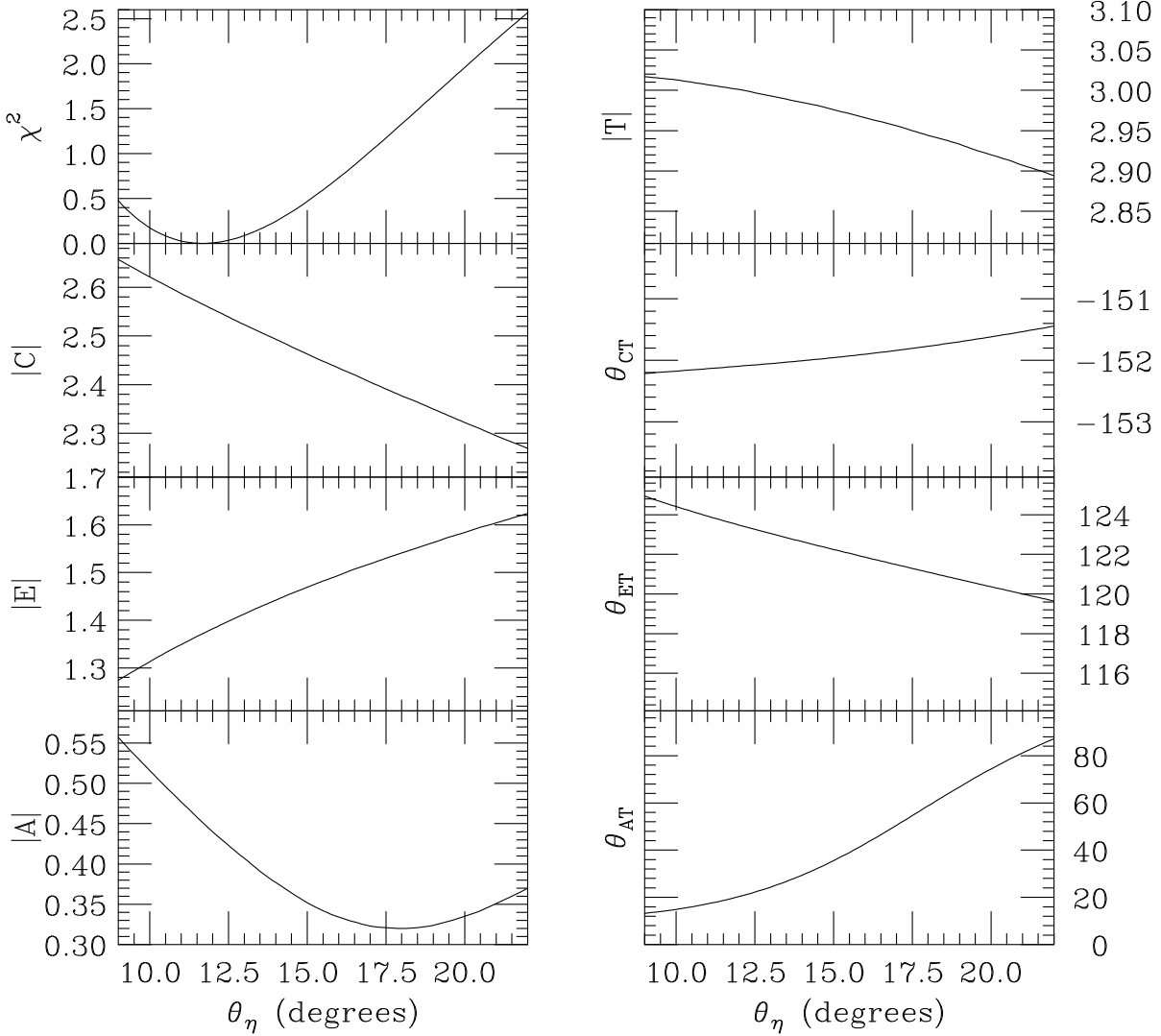}}
\caption{Behavior of $\chi^2$ and Cabibbo-favored decay amplitudes and relative
phases $(|T|, |C|, \theta_{CT}, |E|, \theta_{ET}, |A|, \theta_{AT})$ as
functions of $\theta_\eta$ for the $|T| > |C|$ solution.
\label{fig:bigt}}
\end{figure}

Table \ref{tab:CF2} lists the corresponding representations of the
Cabibbo-favored decay amplitudes as functions of $\thet$. In Fig.\
\ref{fig:bigt} we show the variation of $\chi^2$ and other parameters as
functions of $\thet$ over the range $\thet = 9^\circ - 22^\circ$. The top
left-hand plot shows that $\chi^2$ increases as we move away from the
value $\thet = 11.7^\circ$. Over this range the amplitudes and relative
phases show only a slight change ($< 50\%$) in value as observed in the
other panels of the same Figure, except for the relative phase between
$T$ and $A$. This relative phase ($\theta_{AT}$) varies over a wider
range ($12^\circ$ to $88^\circ$), as shown in the bottom right-hand
plot of Fig.\ \ref{fig:bigt}.

\begin{figure}
\begin{center}
\includegraphics[width=0.7\textwidth]{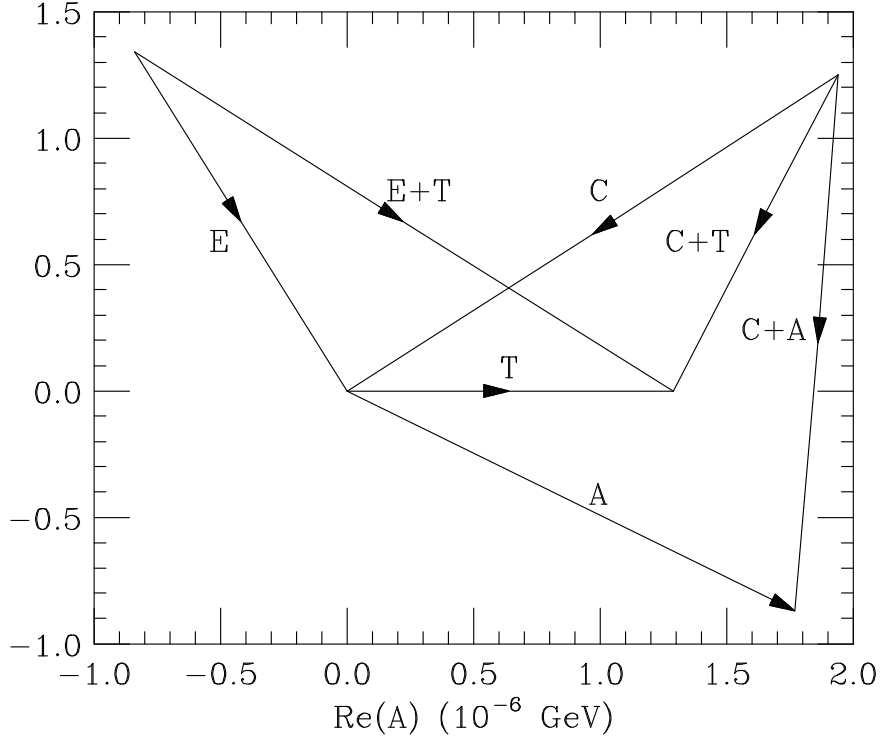}
\end{center}
\caption{Construction of Cabibbo-favored amplitudes from observed processes
using a least $\chi^2$-fit. The sides $C+T$, $C+A$, and $E+T$ correspond to
measured processes; the magnitudes of other amplitudes listed in Table
\ref{tab:CF1} are also needed to specify $T$, $C$, $E$, and $A$. These
figures correspond to the $|T| < |C|$ solution with $\chi^2 = 0$ and
$\thet = 18.9^\circ$.
\label{fig:CFA2}}
\end{figure}

\begin{figure}
\mbox{\includegraphics[width=0.96\textwidth]{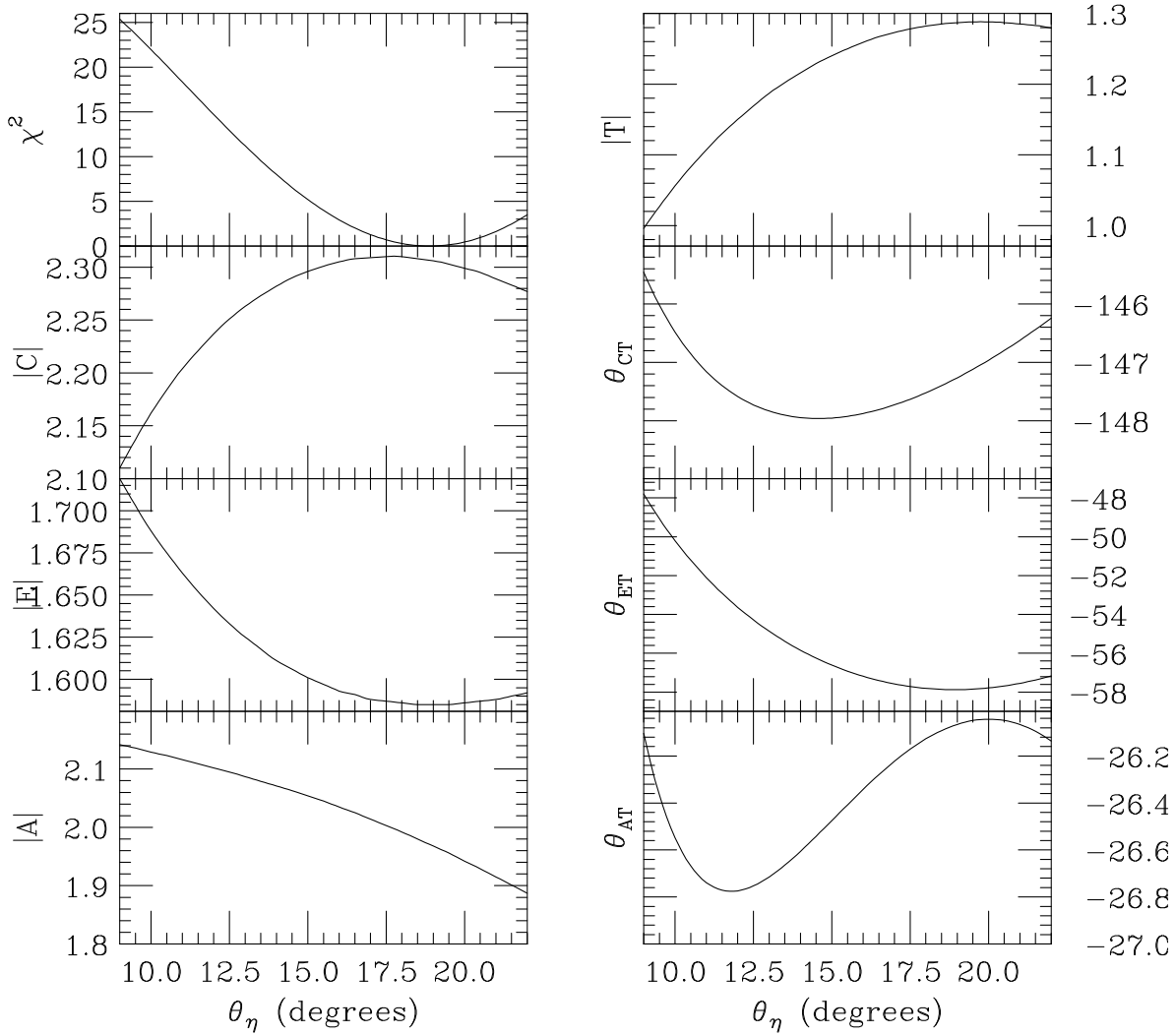}}
\caption{Behavior of $\chi^2$ and Cabibbo-favored decay amplitudes and relative
phases $(|T|, |C|, \theta_{CT}, |E|, \theta_{ET}, |A|, \theta_{AT})$ as
functions of $\theta_\eta$ for the $|T| < |C|$ solution.
\label{fig:smallt}}
\end{figure}

Another set of solutions, but one with $|T| < |C|$, was also found in the
process of minimizing $\chi^2$ as a function of $\theta_\eta$. This branch has
$\chi^2 = 0$ at $\thet = 18.9^\circ$.  Fig.\ \ref{fig:CFA2} shows the Argand
diagram plot for the corresponding amplitudes.  Fig.\ \ref{fig:smallt} shows
the variation of $\chi^2$ and other parameters as functions of $\thet$.  In
this case the point at $\chi^2 = 0$ lies very close to the frequently-quoted
\cite{etamix} value of $19.5^\circ$. The amplitudes and relative phases of
$T,~C,~E$, and $A$ (including the relative phase between $T$ and $A$) show only
slight variations over the range of values considered for $\thet$ ($9^\circ -
22^\circ$.) However the value of $\chi^2$ in the present case shows a much
more rapid increase with a change in $\thet$ as compared to the $|T| > |C|$
case mentioned earlier. The maximum contribution to the $\chi^2$ comes from
the process $D^0 \to \ol{K}^0\eta$ in the $|T|<|C|$ case. To understand this
better let us look at the representation for this amplitude (listed in Table
\ref{tab:CF2}). The part of this representation that depends on $E$ is
proportional to $\sin(\thet-\phi_2)$ which takes the value zero at
$\thet = \phi_2 = 19.5^\circ$. Hence over the range of $\thet$ values we
consider this term doesn't undergo considerable change. The other term
in this representation that depends on $C$ is proportional to
$\sin(\thet+\phi_1)$ which is of the order of unity in this range of $\thet$
values. Now the $|C|$ vs.\ $\thet$ plot in Fig. \ref{fig:smallt} shows that
as we decrease $\thet$, $|C|$ decreases so that the amplitude of the
$D^0 \to \ol{K}^0\eta$ process decreases. This leads to a $3 - 4 \sigma$
variation in the branching ratio and hence a high $\chi^2$
contribution. In the $|T| > |C|$ case, however, Fig. \ref{fig:bigt} shows an
increase in $|C|$ as we reduce $\thet$ from $19.5^\circ$ to $9^\circ$.
This leads to a reasonably stable value of the branching ratio and hence a
very small contribution to the relevant $\chi^2$.

\section{SINGLY-CABIBBO-SUPPRESSED DECAYS}

\subsection{SCS decays involving pions and kaons}

Assuming the relative hierarchy between Cabibbo-favored and singly-%
Cabibbo-suppressed decay amplitudes described in Sec.\ I, and using $\lambda =
\tan\theta_C = 0.2317$ we find, in units of $10^{-7}$ GeV:
\bea
T^\p &=& 6.96;\\
C^\p &=& -5.25 - 2.78\im;\\
E^\p &=& -1.76 + 2.65\im;\\
A^\p &=& 0.99 + 0.34\im.
\eea
where we have considered the $|T| > |C|$ solution for $\thet = 11.7
^\circ$. These amplitudes may then be used to predict the branching
ratios for singly-Cabibbo-suppressed (SCS) $D$ decays. In Table
\ref{tab:SCS} we summarize the measured and predicted amplitudes of
SCS decays to pions and kaons. As was noted in \cite{Bhattacharya:2008ss},
we predict $\cb(D^0\to\pi^+\pi^-)$ larger than observed and
$\cb(D^0\to K^+K^-)$ smaller than observed. This deviation from flavor
$SU(3)$ symmetry is due at least in part to the ratios of decay constants
$f_K/f_\pi = 1.2$ and form factors $f_+(D\to K)/f_+(D\to\pi) > 1$. Other
predictions for singly-Cabibbo-suppressed decays involving pions and kaons
are consistent with those in Ref.\ \cite{Bhattacharya:2008ss}. Table
\ref{tab:SCS} also includes branching ratios predicted using the $|T|
< |C|$ ($\thet = 18.9^\circ$) solution that was obtained in the previous
section. However the predictions for the branching ratios of $D^+\to
K^+\ol{K}^0$ and $D^+_s\to\pi^+K^0$, in this case are an order of
magnitude smaller than their measured values.

\subsection{SCS decays involving $\eta$, $\eta^\p$}

In Table \ref{tab:SCSeta} we quote the branching ratios and extracted
amplitudes for singly-Cabibbo-suppressed D-meson decays involving $\eta$ and
$\eta^\p$ as reported in \cite{Mendez:2009}. The values of $C^\p$ and $E^\p$
obtained in the previous section may be used to determine the relevant parts of
the amplitudes for $D^0$ decays involving $\eta$ and $\eta^\p$.  Additional
``disconnected'' flavor-singlet diagrams $SE^\p$ and $SA^\p$
\cite{Chiang:2002mr} are required. In Table \ref{tab:SCSth} we show the
representations of the above amplitudes as a function of $\thet$ using Eqs.\
(\ref{eqn:a1},\ref{eqn:a2}).

\begin{table}
\caption{Branching ratios and invariant amplitudes for
singly-Cabibbo-suppressed decays of charmed mesons to pions and kaons.
\label{tab:SCS}}
\begin{center}
\begin{tabular}{c l c c c c c c}
\hline \hline
Meson & Decay & $\cb$\cite{Mendez:2009} & $p^*$ & $|{\cal A}|$  & Rep.\
 & \multicolumn{2}{c}{Predicted $\cb$ ($10^{-3}$)} \\
 & mode & ($10^{-3}$) & (MeV) &$(10^{-7} GeV)$& & $|T|<|C|$
 & $|T|>|C|$ \\ \hline
$D^0$ &$\pi^+ \pi^-$   &$1.45\pm0.05$& 921.9 & $4.70\pm0.08$ &$-(T^\p+E^\p)$
 & 2.24 & 2.24  \\
 &$\pi^0 \pi^0$   &$0.81\pm0.05$& 922.6 & $3.51\pm0.11$ &$-(C^\p-E^\p)/\s$
 & 1.36 & 1.35  \\
 &$K^+   K^-$     &$4.07\pm0.10$& 791.0 & $8.49\pm0.10$ &$ (T^\p+E^\p)$
 & 1.92 & 1.93  \\
 &$K^0 \ol{K}^0$&$0.32\pm0.02$& 788.5 & $2.39\pm0.14$ & $0$ & 0 & 0 \\ \hline
$D^+$ &$\pi^+ \pi^0$ &$1.18\pm0.06$& 924.7 & $2.66\pm0.07$ &$-(T^\p+C^\p)/\s$
 & 0.88 & 0.89  \\
 &$K^+   \ol{K}^0$&$6.12\pm0.22$& 792.6 & $6.55\pm0.12$ &$(T^\p-A^\p)$
 & 0.73 & 6.15  \\ \hline
$D^+_s$&$\pi^+ K^0$ &$2.52\pm0.27$& 915.7 & $5.94\pm0.32$ &$-(T^\p-A^\p)$
 & 0.37 & 3.08  \\
 &$\pi^0 K^+$ &$0.62\pm0.23$& 917.1 & $2.94\pm0.55$ &$-(C^\p+A^\p)/\s$
 & 0.86 & 0.85  \\ \hline \hline
\end{tabular}
\end{center}
\end{table}

\begin{table}
\caption{Branching ratios and amplitudes for singly-Cabibbo-suppressed decays
of $D^0$, $D^+$ and $D^+_s$ involving $\eta$ and $\eta^\p$. Here we use the
representations quoted in Eqn. \ref{eqn:b}.
\label{tab:SCSeta}}
\begin{center}
\begin{tabular}{c l c c c c} \hline \hline
Meson & Decay &$\cb$\cite{Mendez:2009}& $p^*$ & $|{\cal A}|$  & Rep.\\
      & mode  &  ($10^{-4}$)          & (MeV) &$(10^{-7} GeV)$&     \\ \hline
$D^0$ &$\pi^0 \eta$   &$6.80\pm0.70$ & 846.2 & $3.36\pm0.17$
 &$-\frac{1}{\sx}(2E^\p-C^\p+SE^\p)$      \\
 &$\pi^0 \eta^\p$&$9.10\pm1.27$ & 678.0 & $4.34\pm0.30$
 &$\frac{1}{\st}(E^\p+C^\p+2SE^\p)$       \\
 &$\eta  \eta$   & $16.7\pm1.8$ & 754.7 & $5.57\pm0.30$
 &$\frac{2\s}{3}(C^\p+SE^\p)$             \\
 &$\eta  \eta^\p$& $10.5\pm2.6$ & 536.9 & $5.24\pm0.65$
 &$-\frac{1}{3\s}(C^\p+6E^\p+7SE^\p)$     \\ \hline
$D^+$ &$\pi^+ \eta$   & $35.4\pm2.1$ & 848.4 & $4.82\pm0.14$
 &$\frac{1}{\st}(T^\p+2C^\p+2A^\p+SA^\p)$ \\
 &$\pi^+ \eta^\p$& $46.8\pm3.0$ & 680.5 & $6.18\pm0.20$
 &$-\frac{1}{\sx}(T^\p-C^\p+2A^\p+4SA^\p)$\\ \hline
$D^+_s$&$K^+   \eta$   & $17.6\pm3.6$ & 835.0 & $5.20\pm0.53$
 &$\frac{1}{\st}(T^\p+2C^\p-SA^\p)$       \\
 &$K^+   \eta^\p$& $18.0\pm5.0$ & 646.1 & $5.98\pm0.83$
 &$\frac{1}{\sx}(2T^\p+C^\p+3A^\p+4SA^\p)$\\ \hline \hline
\end{tabular}
\end{center}
\end{table}

\begin{table}
\caption{Representations for singly-Cabibbo-suppressed decays of $D^0$, $D^+$
and $D^+_s$ involving $\eta$ and $\eta^\p$ for an arbitrary $\eta - \eta^\p$
mixing angle $\thet$. Here we use the representations quoted in Eqs.\
(\ref{eqn:a1}, \ref{eqn:a2}). ($\phi_1 = 45^\circ - \frac{\phi_2}{2}$
and $\phi_2 = 19.5^\circ$.)
\label{tab:SCSth}}
\begin{center}
\begin{tabular}{c l c} \hline \hline
Meson  & Decay & Representation                                 \\
       & mode  &                                                \\ \hline
$D^0$  &$\pi^0 \eta$
&$\frac{C^\p}{\s}\cos(\thet+\phi_1)-E^\p~\sin(\thet+\phi_1)
-\frac{\st~SE^\p}{\s}~\sin\thet$ \\
 &$\pi^0 \eta^\p$ &$\frac{C^\p}{\s}\sin(\thet+\phi_1)+E^\p~\cos(\thet+\phi_1)
 +\frac{\st~SE^\p}{\s}~\cos\thet$ \\
 &$\eta  \eta$ &$\frac{\st~C^\p}{\s}\cos\thet~\sin(\thet+\phi_1)
 -\frac{3~E^\p}{\s}\cos\thet~\cos(\thet+2\phi_1)
 +\frac{3~SE^\p}{2}\sin(2\thet)$ \\
 &$\eta  \eta^\p$ &$-\frac{\st~C^\p}{2}\cos(2\thet+\phi_1)
 +\frac{3~E^\p}{2}\sin(2\thet+2\phi_1)-\frac{3~SE^\p} {\s}\cos(2\thet)$ \\
 \hline $D^+$  &$\pi^+ \eta$
 &$~~\frac{T^\p}{\s}\sin(\thet+\phi_1)+\frac{\st~C^\p}{\s}\cos\thet
 +\s~A^\p~\sin(\thet+\phi_1) +\st~SA^\p~\sin\thet$ \\
 &$\pi^+ \eta^\p$ &$-\frac{T^\p}{\s}\cos(\thet+\phi_1)
 +\frac{\st~C^\p}{\s}\sin\thet-\s~A^\p~\cos(\thet+\phi_1)
 -\st~SA^\p~\cos\thet$ \\ \hline
$D^+_s$&$K^+ \eta$ &$~T^\p~\cos(\thet+\phi_1)+\frac{\st~C^\p}{\s}\cos\thet~
 +~\frac{\st~A^\p}{\s}\cos(\thet+2\phi_1)
 -\st~SA^\p~\sin\thet$ \\
 &$K^+   \eta^\p$ &$~T^\p~\sin(\thet+\phi_1)+\frac{\st~C^\p}{\s}\sin\thet~
 -~\frac{\st~A^\p}{\s}~\sin(\thet+2\phi_1) +\st~SA^\p~\cos\thet$ \\
\hline \hline
\end{tabular}
\end{center}
\end{table}

In Tables \ref{tab:SEp19.5} ($\thet = 19.5^\circ$) and \ref{tab:SEp11.7}
($\thet = 11.7^\circ$) we write the amplitudes so that the coefficient of
$SE^\p$ or $SA^\p$ is always one. As explained in \cite{Bhattacharya:2008ss,
Rosner:1999} this information may be used to determine $SE^\p$ and $SA^\p$
using a plotting technique.

\begin{figure}
\mbox{\includegraphics[width=0.48\textwidth]{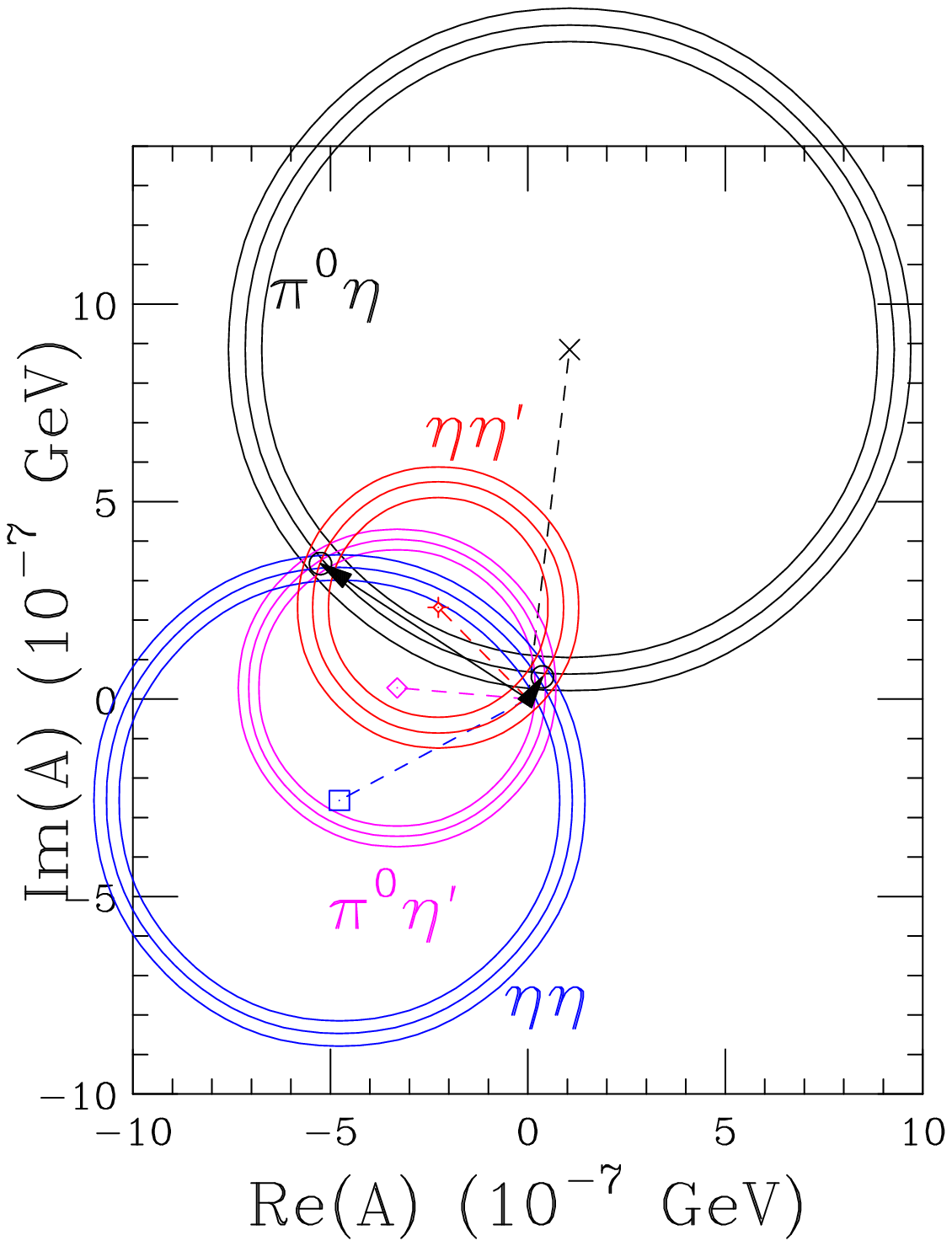}
      \includegraphics[width=0.50\textwidth]{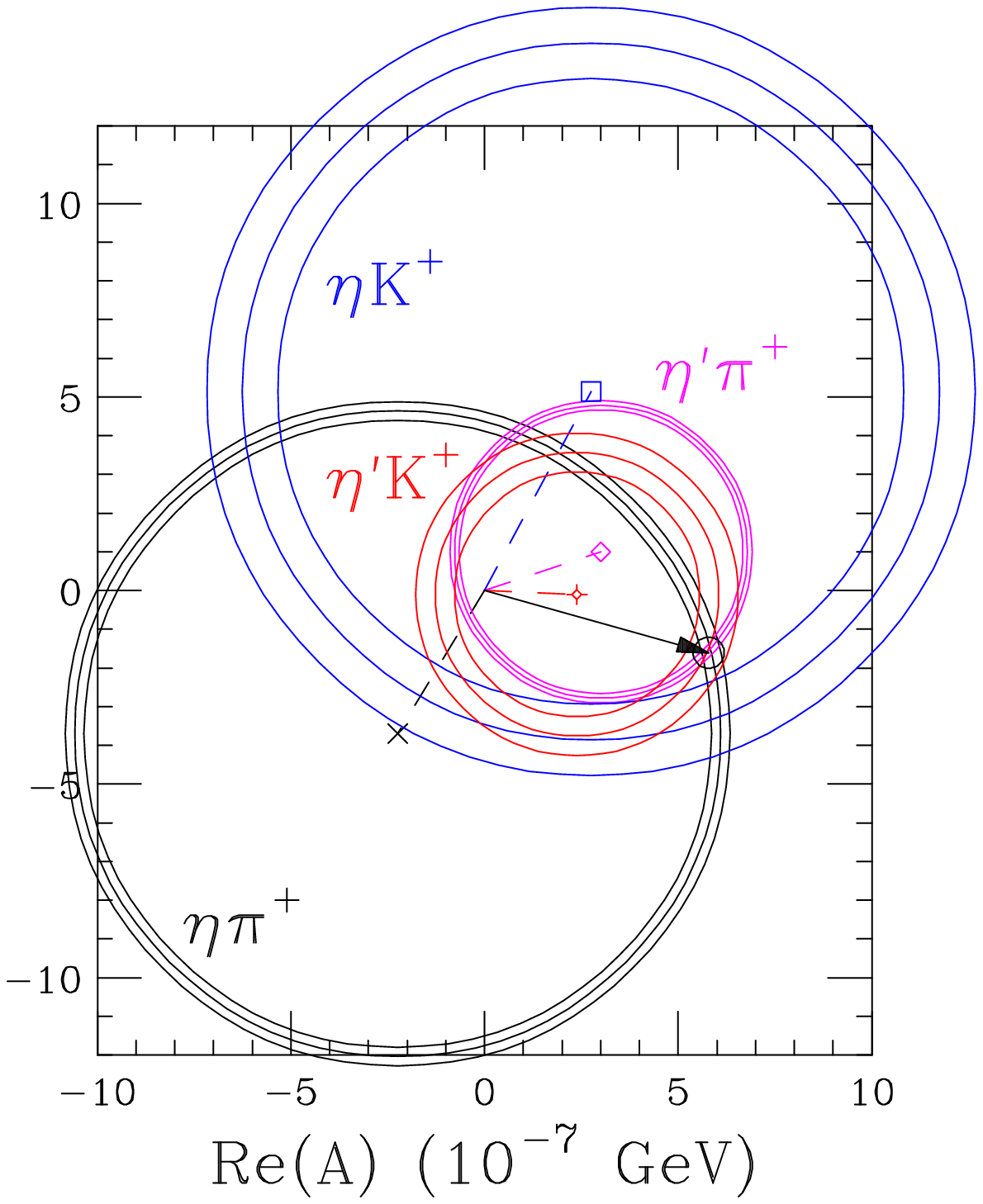}}
\caption{Determination of the disconnected singlet annihilation amplitudes
$SE'$ (left) and $SA'$ (right) from SCS charmed meson decays involving $\eta$
and $\eta'$ in the solutions with $|T| > |C|$ and $\thet = 19.5^\circ$.  Left:
$D^0$ decays to final states as shown; right:  $D^+$ or $D_s^+$ decays to final
states as shown.  The small black circles show the solution regions.  Arrows
pointing to them denote the complex amplitudes $-SE^\p$ (left) and $-SA^\p$
(right).
\label{fig:SEpSAp19}}
\end{figure}

\begin{figure}
\mbox{\includegraphics[width=0.52\textwidth]{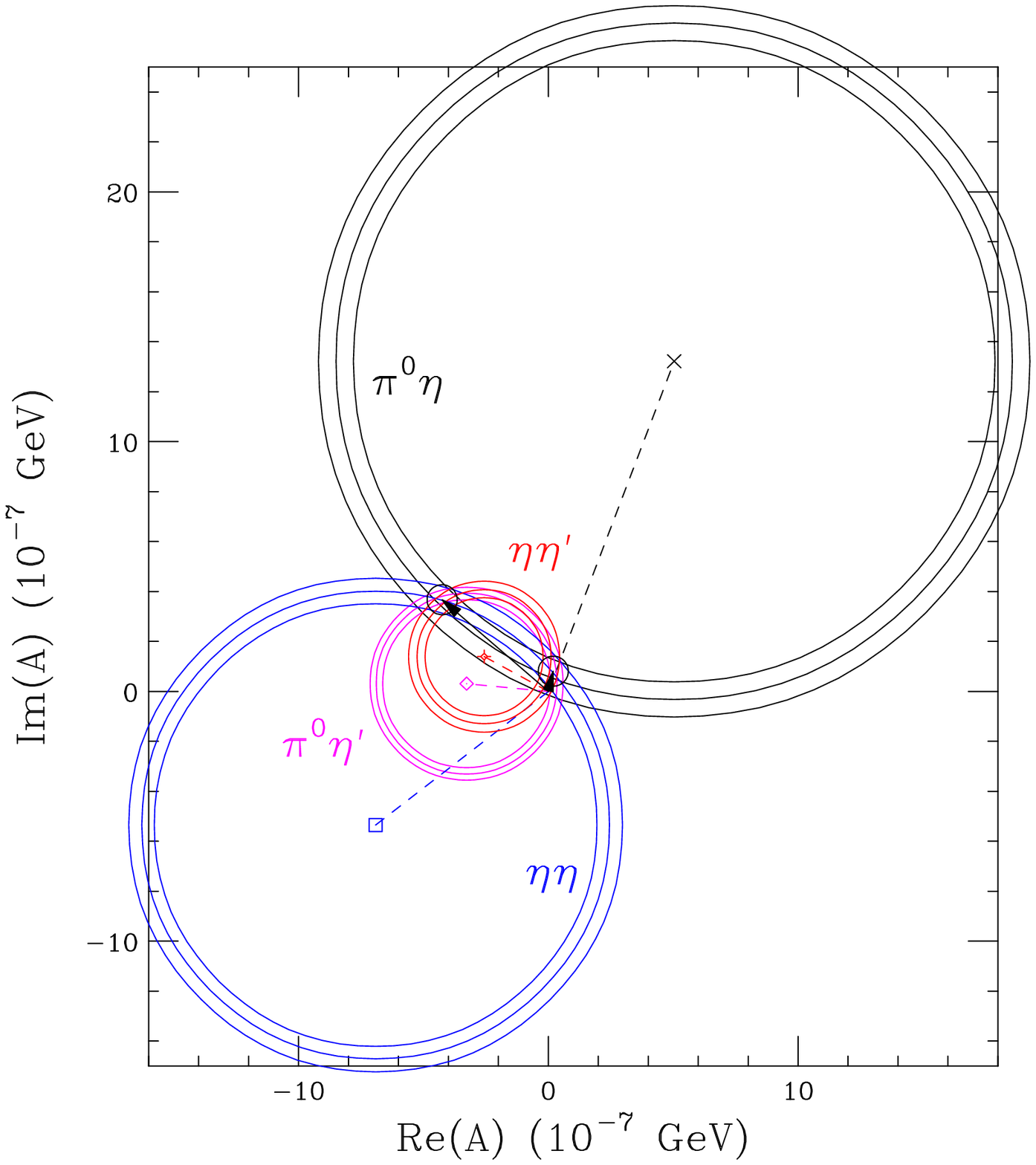}
      \includegraphics[width=0.46\textwidth]{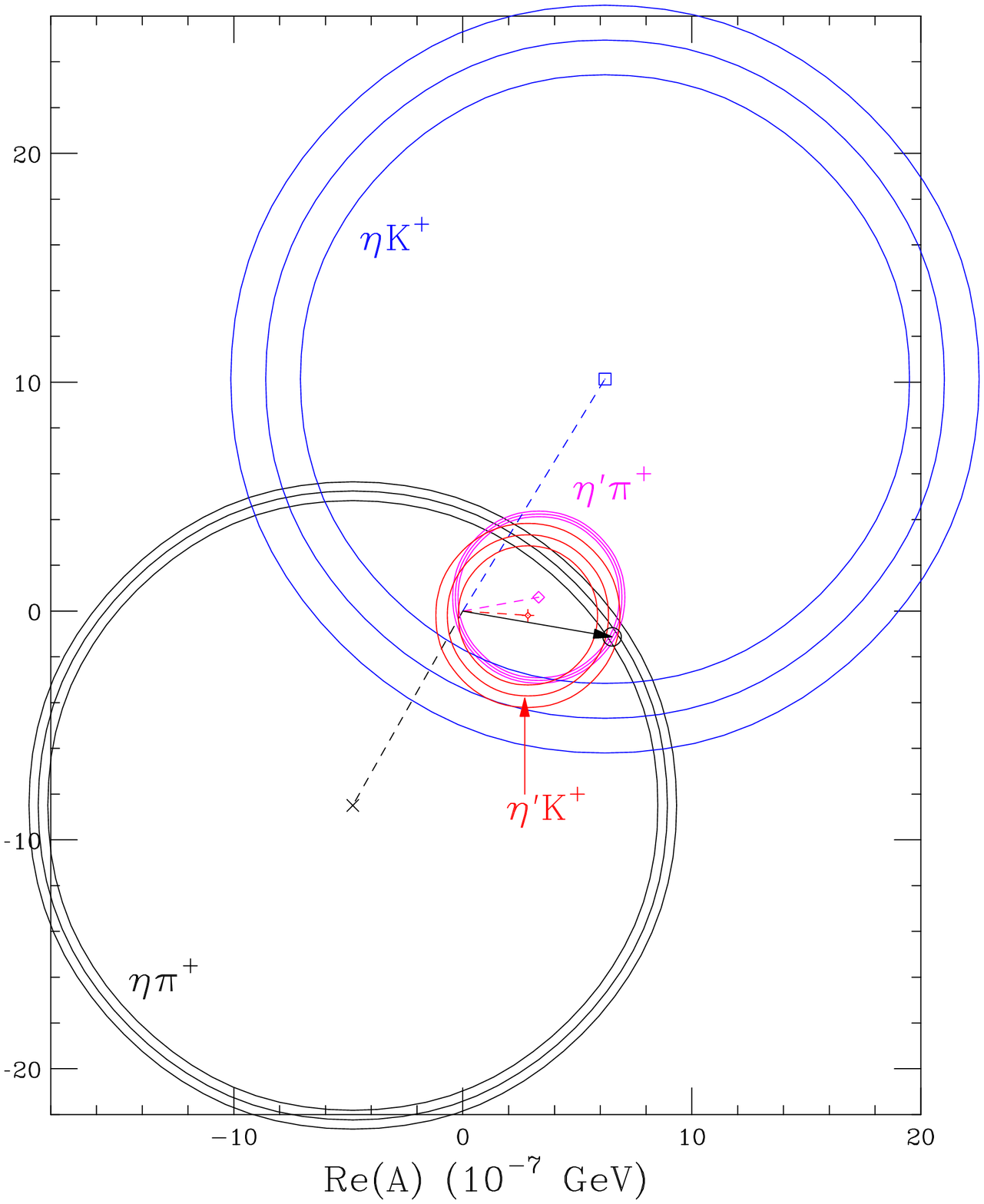}}
\caption{Same as Fig.\ 5 except $\thet = 11.7^\circ$.
\label{fig:SEpSAp}}
\end{figure}

In Fig.\ \ref{fig:SEpSAp19} we show the construction technique to find the
singlet amplitudes for $\thet = 19.5^\circ$. A $\chi^2$ minimization fit was
used to obtain the amplitudes. We find both a small and a large solution for
$SE^\p$ with a small value for $\chi^2$ of around 0.215 for both solutions.  In
units of $10^{-7}$ GeV, they are $SE^\p = -(0.37\pm0.20)-(0.56\pm0.31)\im$ and
$SE^\p=(5.25\pm0.29)-(3.43 \pm0.21)\im$. The second is less likely, as it has
too large an amplitude when compared to the connected amplitudes. For $SA^\p$,
only a large solution ($SA^\p = -(5.79\pm0.14) + (1.61\pm0.17)\im$ with $\chi^2
= 2.6$) is found using a $\chi^2$ minimization fit. A similar exercise was
carried out for $\thet = 11.7^ \circ$. The corresponding construction is shown
in Fig.\ \ref{fig:SEpSAp}. In this case we once again find two solutions for
$SE^\p$ (one large and one small in magnitude) and only one solution for
$SA^\p$. These are $SE^\p=-(0.20\pm0.20)-(0.81\pm0.48)\im$ ($\chi^2 = 0.5$),
$SE^\p=(4.25^{+0.37}_{-0.47})-(3.67\pm0.21)\im$ ($\chi^2 = 0.5$) and $SA^\p =
-(6.52\pm0.13) + (1.13 \pm0.24)\im$ ($\chi^2 = 4.9$.) Qualitatively similar
conclusions were reached in \cite{Bhattacharya:2008ss} where we used only the
value $\thet = 19.5^\circ$.

\begin{table}
\caption{Real and imaginary parts of amplitudes for singly-Cabibbo-suppressed
decays of $D^0$, $D^+$ and $D^+_s$ involving $\eta$ and $\eta^\p$ ($\thet =
19.5^\circ$), in units of $10^{-7}$ GeV, used to generate plots in Fig.\
\ref{fig:SEpSAp19}.
\label{tab:SEp19.5}}
\begin{center}
\begin{tabular}{c c c c c}
\hline \hline
Amplitude($\ca$ & Expression & Re  & Im  & $|\ca_{\rm exp}|$\\ \hline
$-\sx\ca(D^0\to\pi^0\eta)$ & $2E^\p-C^\p+SE^\p$ & 1.06& 8.85&$8.22\pm0.42$\\
$\frac{\st}{2}\ca(D^0\to\pi^0\eta^\p)$ &$\frac{1}{2}(C^\p+E^\p)+SE^\p$
 &-3.31& 0.28&$3.76\pm0.26$\\
$\frac{3}{2\s}\ca(D^0\to\eta\eta)$ &$C^\p+SE^\p$ & -4.77&-2.57&$5.91\pm0.32$\\
$-\frac{3\s}{7}\ca(D^0\to\eta\eta^\p)$ &$\frac{1}{7}(C^\p+6E^\p)+SE^\p$
 &-2.27& 2.32&$3.17\pm0.39$\\ \hline
$\st\ca(D^+\to\pi^+\eta)$ &$T^\p+2C^\p+2A^\p+SA^\p$&-2.24&-3.70&$8.34\pm0.25$\\
$-\frac{\sx}{4}\ca(D^+\to\pi^+\eta^\p)$&$\frac{1}{4}(T^\p-C^\p+2A^\p)+SA^\p$
 & 3.01& 1.00&$3.79\pm0.12$\\ \hline
$-\st\ca(D^+_s\to\eta K^+)$ &$-(T^\p+2C^\p)+SA^\p$ & 2.75& 5.14&$9.00\pm0.92$\\
$\frac{\sx}{4}\ca(D^+_s\to\eta^\p K^+)$&$\frac{1}{4}(2T^\p+C^\p+3A^\p)+SA^\p$
 & 2.39&-0.10&$3.66\pm0.51$\\ \hline \hline
\end{tabular}
\end{center}
\end{table}

\begin{table}
\caption{Real and imaginary parts of amplitudes for singly-Cabibbo-suppressed
decays of $D^0$, $D^+$ and $D^+_s$ involving $\eta$ and $\eta^\p$($\thet =
11.7^\circ$), in units of $10^{-7}$ GeV, used to generate plots in Fig.\
\ref{fig:SEpSAp}.
\label{tab:SEp11.7}}
\begin{center}
\begin{tabular}{c c c c c c c c c c}
\hline \hline
Amplitude & \multicolumn{6}{c}{Coefficients in the expression} & Re  & Im  & $|\ca_{\rm exp}|$\\
          & $T^\p$ & $C^\p$ & $E^\p$ & $A^\p$ & $SE^\p$ & $SA^\p$ & & & \\ \hline
$-4.03~\ca(D^0\to\pi^0\eta)$      &    0&-1.95& 2.95&    0&1&0&~5.04&13.22&13.54$\pm$0.70  \\
$~~~0.83~\ca(D^0\to\pi^0\eta^\p)$ &    0& 0.43& 0.57&    0&1&0&-3.27&~0.31&~3.62$\pm$0.25  \\
$~1.68~\ca(D^0\to\eta\eta)$       &    0& 1.47&-0.47&    0&1&0&-6.91&~5.35&~9.37$\pm$0.51  \\
$-0.51~\ca(D^0\to\eta\eta^\p)$    &    0& 0.23& 0.77&    0&1&0&-2.57&~1.39&~2.69$\pm$0.33  \\ \hline
$~~~~2.85~\ca(D^+\to\pi^+\eta)$   & 1.47& 3.42&    0& 2.95&0&1&-4.80&-8.50&13.74$\pm$0.41  \\
$~~~~0.59~\ca(D^+\to\pi^+\eta^\p)$&-0.28& 0.15&    0&-0.57&0&1&~3.31&~0.60&~3.65$\pm$0.12  \\ \hline
$-2.85~\ca(D^+_s\to\eta\eta)$     &-1.95&-3.42&    0& 1.81&0&1&~6.21&10.13&14.83$\pm$1.52  \\
$~~~0.59~\ca(D^+_s\to\eta\eta^\p)$   & 2.84&-0.19&    0& 0.62&0&1& 2.84&-0.19&~3.52$\pm$0.49  \\ \hline \hline
\end{tabular}
\end{center}
\end{table}

\subsection{Sum rules for $D^0 \to (\pi^0\pi^0, \pi^0\eta, \eta\eta,
\pi^0\eta^\p, \eta\eta^\p)$}

Let us consider the singly-Cabibbo-suppressed decays of the $D^0$, where the
final pseudoscalars are $\pi^0$, $\eta$, or $\eta^\p$. In Tables
\ref{tab:SCS} and \ref{tab:SCSeta} we list the flavor topology representations
for all such decays.  It is interesting to note that there are 5 such
amplitudes all of which depend only on $C^\p$, $E^\p$ and $SE^\p$. This means
we can algebraically relate two or more of these amplitudes through sum rules,
such as
\bea
\s~\ca(D^0\to\pi^0\pi^0) + \st~\sin\thet~\ca(D^0\to\pi^0\eta^\p) +
\st~\cos\thet~\ca(D^0\to\pi^0\eta) = 0~, \\
(1-3~\sin^2\thet)~\ca(D^0\to\pi^0\pi^0) + \ca(D^0\to\eta\eta) - \sx~\sin
\thet~\ca(D^0\to\pi^0\eta^\p) = 0~, \\
(2~\cos^2\thet\cos(2~\thet)+\sin^2(2~\thet))~\ca(D^0\to\pi^0\pi^0)
+ 2~\cos(2~\thet)~\ca(D^0\to\eta\eta) \nonumber \\
+ \s~\sin(2~\thet)~\ca(D^0\to\eta\eta^\p) = 0~.
\eea
A sum rule relating three amplitudes can be represented by a triangle whose
sides are the magnitudes of the corresponding amplitudes.  As in
\cite{Bhattacharya:2008ss}, these triangles have angles not equal to zero or
$180^\circ$, showing that the amplitudes have non-zero relative strong phases.

The sum rule in \cite{Bhattacharya:2008ss} relating squares of magnitudes of
amplitudes,
\bea
8|\ca(D^0\to\pi^0\eta^\p)|^2 + 16|\ca(D^0\to\pi^0\pi^0)|^2 \nonumber \\
= 16|\ca(D^0\to\pi^0\eta)|^2 +  9|\ca(D^0\to\eta\eta)|^2~,
\eea
may be re-evaluated using the data from Tables \ref{tab:SCS} and
\ref{tab:SCSeta}.  We find, in units of $10^{-14}$ GeV$^2$,
\bea
8|\ca(D^0\to\pi^0\eta^\p)|^2 + 16|\ca(D^0\to\pi^0\pi^0)|^2 = 348\pm24 \\
16|\ca(D^0\to\pi^0\eta)|^2 +  9|\ca(D^0\to\eta\eta)|^2 = 460\pm35
\eea
The deviation from equality by about $2.6\sigma$ indicates the degree of
flavor-SU(3) breaking.

The above sum rule refers to the special case of $\thet = 19.5^\circ$.
A more general sum rule, valid for arbitrary values of $\thet$, is
\bea
81\sin^2\thet\cos^2\thet|\ca(D^0\to\pi^0\eta^\p)|^2
+ 27\cos^2\thet(3\cos^2\thet-2)|\ca(D^0\to\pi^0\pi^0)|^2 \nonumber \\
= 27\cos^2\thet(3\cos^2\thet-2)|\ca(D^0\to\pi^0\eta)|^2
+ 9|\ca(D^0\to\eta\eta)|^2~,
\eea
where we have kept the normalization of the previous sum rule.  The left-hand
side of this sum rule, in units of $10^{-14}$ GeV$^2$, is $340 \pm 19$, while
the right-hand side is $535 \pm 40$.  Here the discrepancy rises to $4.4
\sigma$.  Allowing $\thet$ to be a free parameter does not improve the
description of SCS decays.

\section{DOUBLY-CABIBBO-SUPPRESSED DECAYS}

\begin{figure}
\mbox{\includegraphics[width=0.96\textwidth]{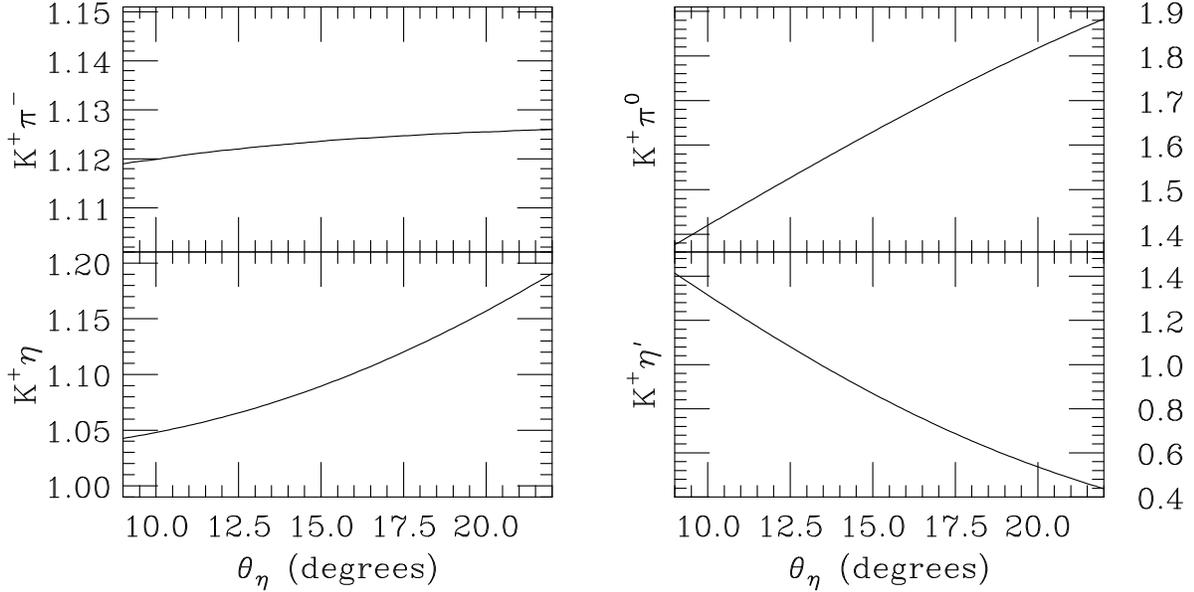}}
\caption{Branching ratios in units of $10^{-4}$ of doubly-Cabibbo-suppressed
decays $D^0\to K^+\pi^-, D^0\to K^0(\pi^0,\eta,\eta^\p), D^+\to K^+(\eta,
\eta^\p)$ as functions of $\theta_\eta$ for the $|T| > |C|$ solution.
\label{fig:ws}}
\end{figure}

Table \ref{tab:DCS} contains a list of doubly-Cabibbo-suppressed decay
branching ratios, amplitudes and their representation in terms of $\ttl$,
$\tc$, $\te$, and $\ta$. The magnitudes of these are obtained by multiplying
the corresponding Cabibbo-favored amplitudes listed in Sec. III by
$- \lambda^2$. Using $\lambda = 0.2317$ we find
$\cb(D^0 \to K^+ \pi^-) = 1.12 \times 10^{-4}$ and $\cb(D^+ \to K^+ (\pi^0,
\eta, \eta^\p) = (1.49,1.06,1.16) \times 10^{-4}$,
where we have used $\thet = 11.7^\circ$
and the $|T| > |C|$ solution.  While the experimental branching ratio for
$D^0 \to K^+ \pi^-$ remains $29\%$ above the prediction from flavor $SU(3)$,
the measured branching ratio for $D^+ \to K^+ \pi^0$ matches the predicted
value to around $1 \sigma$.
In Fig.\ \ref{fig:ws} we plot the branching ratios of the predicted
doubly-Cabibbo-suppressed decays $D^0\to K^-\pi^+$ and
$D^+\to K^+(\pi^0, \eta, \eta^\p)$, which are ones for which 
measurements or upper bounds exist, as functions of $\thet$.

\begin{table}
\caption{Branching ratios and amplitudes for doubly Cabibbo-suppressed decays
of $D^0$, $D^+$ and $D^+_s$.
\label{tab:DCS}}
\begin{center}
\begin{tabular}{c l c c c c} \hline \hline
Meson & Decay &$\cb$\cite{Mendez:2009}& $p^*$ & Representation
 & Predicted $\cb$      \\
 & mode  &  ($10^{-4}$)      & (MeV) &      & $\thet = 11.7^\circ$ \\ \hline
 $D^0$ &$K^+ \pi^-$  &$1.45\pm0.04$&861.1&$\ttl+\te$       & 1.12  \\
       &$K^0 \pi^0$  &             &860.4&$(\tc-\te)/\s$   & 0.69  \\
       &$K^0 \eta$   &             &771.9&$~~~\frac{\tc}{\s}\sin(\thet+\phi_1)
       -\frac{\st\te}{\s}\cos(\thet+2\phi_1)$  & 0.28     \\
       &$K^0 \eta^\p$&             &564.9&$-\frac{\tc}{\s}\cos(\thet+\phi_1)
       -\frac{\st\te}{\s}\sin(\thet+2\phi_1)$& 0.55      \\ \hline
 $D^+$ &$K^0 \pi^+$  &             &862.6&$\tc+\ta$        & 2.01  \\
       &$K^+ \pi^0$  &$1.72\pm0.19$&864.0&$(\ttl-\ta)/\s$  & 1.49  \\
       &$K^+ \eta$   &$<1.3$       &775.8&$-\frac{\ttl}{\s}\sin(\thet+\phi_1)
       -\frac{\st\ta}{\s}\cos(\thet+2\phi_1)$&  1.06 \\
       &$K^+ \eta^\p$&$<1.8$       &570.8&$~~~\frac{\ttl}{\s}\cos(\thet+\phi_1)
       +\frac{\st\ta}{\s}\sin(\thet+2\phi_1)$&  1.16 \\ \hline
$D^+_s$&$K^0 K^+$    &             &850.3&$\ttl+\tc$       & 0.38  \\
\hline \hline
\end{tabular}
\end{center}
\end{table}

As described in \cite{Bhattacharya:2008ss} in general the decays
$D \to (K\pi, \ol{K}\pi)$ are related to each other by either a simple U-spin
interchange $s\leftrightarrow d$, or by interchanging ``Tree" and
``Annihilation" amplitudes.  In the former case, the extent of $SU(3)$ breaking
in the prediction for the decay asymmetry is expected to be very small.

In $D^0$ decays the asymmetry is predicted as a consquence of U-spin:
\bea
R(D^0) &\equiv& \frac{\Gamma(D^0\to K_S\pi^0) - \Gamma(D^0\to K_L\pi^0)}
{\Gamma(D^0\to K_S\pi^0) + \Gamma(D^0\to K_L\pi^0)} \\
&=& 2 \lambda^2 \,=\, 0.107~.
\eea
and is indeed consistent with the observed value \cite{He:2008}
\beq
R(D^0) = 0.108 \pm 0.025 \pm 0.024~.
\eeq
Similarly for the $D^+$ and $D_s^+$ decays one may construct the following
quantities and predict:
\bea
R(D^+) &\equiv& \frac{\Gamma(D^+\to K_S\pi^+) - \Gamma(D^+\to K_L\pi^+)}
{\Gamma(D^+\to K_S\pi^+) + \Gamma(D^+\to K_L\pi^+)} \\
&=& 2 \lambda^2 {\rm Re}\frac{C + A}{T + C}\,=\, - 0.005 \pm 0.013~,\\
R(D_s^+) &\equiv& \frac{\Gamma(D_s^+\to K_S K^+) - \Gamma(D_s^+\to K_L K^+)}
{\Gamma(D_s^+\to K_S K^+) + \Gamma(D_s^+\to K_L K^+)} \\
&=& 2 \lambda^2 {\rm Re}\frac{C + T}{A + C}\,=\, - 0.0022 \pm 0.0087~.
\eea
Here we have made use of the amplitudes obtained from the $\chi^2$ minimum
solution that satisifies $|T| > |C|$ and allows for $\thet = 11.7^\circ$.
These represent similarity with respect to the predictions in Ref.\
\cite{Bhattacharya:2008ss}, $R(D^+) = -0.006^{+0.033}_{-0.028}$ and
$R(D_s^+) = -0.003^{+0.019}_{-0.017}$.  Experimentally only the first
asymmetry is measured \cite{He:2008}:
\beq
R(D^+) = 0.022 \pm 0.016 \pm 0.018~,
\eeq
in agreement with the new prediction just as with the earlier one.

\section{CONCLUSIONS}

We have re-analyzed the decays of charmed mesons to pairs of light
pseudoscalars using flavor SU(3) in the light of new experimental
determinations of branching ratios \cite{Mendez:2009} with experimental
errors often smaller than previous world averages \cite{Amsler:2008}.  The main
difference with respect to a previous analysis \cite{Bhattacharya:2008ss} is
that the ``annihilation'' amplitude $A$, found previously to have a phase
of almost $180^\circ$ with respect to the ``exchange'' amplitude $E$, now
has a phase of $~\sim(100 \pm 10)^\circ$ with respect to $E$ for the preferred
set of amplitudes and phases. While consequences for singly-Cabibbo-suppressed
are qualitatively similar to those in \cite{Bhattacharya:2008ss},
similar decay asymmetries involving $D^+ \to K_{(S,L)} \pi^+$
and $D_s^+ \to K_{(S,L)} \pi^+$ are predicted; both are in good agreement with
the observed value as well as the ones obtained previously.

\section*{ACKNOWLEDGMENTS}
This work was supported in part by the United States Department
of Energy through Grant No.\ DE FG02 90ER40560.

\end{document}